\documentclass{iau}

\usepackage{amsmath}
\usepackage{graphicx}
\usepackage{multirow}
\usepackage{float}

\begin{document}

\lefttitle{Cambridge Author}
\righttitle{Proceedings of the International Astronomical Union: \LaTeX\ Guidelines for~authors}

\jnlPage{1}{7}
\jnlDoiYr{2021}
\doival{10.1017/xxxxx}

\aopheadtitle{Proceedings IAU Symposium}
\editors{C. Sterken,  J. Hearnshaw \&  D. Valls-Gabaud, eds.}

\title{Geomagnetic Storms and Satellite Orbital Decay}

\author{Yoshita Baruah}
\affiliation{Department of Physical Sciences\\
Indian Institute of Science Education and Research Kolkata \\
Mohanpur 741246, West Bengal, India}
\affiliation{Center of Excellence in Space Sciences India\\
Indian Institute of Science Education and Research Kolkata \\
Mohanpur 741246, West Bengal, India}

\begin{abstract}
Energetic events on the Sun, particularly coronal mass ejections and high speed streams, regulate the near Earth space environment and give rise to space weather. A major terrestrial manifestation of such events are geomagnetic storms. A geomagnetic storm results in dissipation of energy from the solar wind into the atmosphere, leading to Joule heating and thermospheric expansion. This has serious consequences on Low Earth Orbit satellite lifetimes. Our work demonstrates the impact of different kinds of geomagnetic storms on satellite orbits. We also briefly discuss about some physical attributes of satellites that can make them prone to higher orbital decay. Our work highlights the importance of monitoring and predicting space weather, and assessing their impacts on space-based human technologies. 
\end{abstract}


\maketitle

\section{Introduction}

Transient energetic events occurring on the Sun -- like solar flares, coronal mass ejections (CMEs), solar energetic particles (SEPs) and high speed streams which form stream interaction regions (SIRs) and corotating interaction regions (CIRs) -- modulate the space environment near Earth. Such activity give rise to space weather and can have serious implications on the performance and reliability of our space and ground based technological infrastructure \citep{Nandy2023,Schrijver2015,Schwenn2006}. This makes space weather assessments, monitoring and predictions extremely important for today's technology-reliant human society \citep{Pal2022, Sinha2022, Roy2023, Oliveira2024}.

This work focuses on the impact of space weather on our space based assets. Earth-directed solar eruptive events like CMEs -- which are often triggered by solar flares and solar filament eruptions -- interact with the terrestrial magnetic field and trigger moderate to severe geomagnetic storms \citep{Pulkkinen2007,Sinha2019,Pal2018,jaswal2024}. CIRs, created when high speed streams emanating from coronal holes interact with the slower ambient solar wind, also drive weak to moderate geomagnetic storms \citep{Tsurutani2006}. These storms are characterised by enhanced particle injection and energy dissipation from the solar wind into the magnetosphere-atmosphere system. A large part of this energy is manifested in the atmosphere as Joule heating \citep{knipp2004}. This causes an upwelling of the thermosphere leading to increased density in the altitude of Low Earth Orbit (LEO) satellites -- resulting in increased drag on them and enhance their orbital decay \citep{Oliveira2017, Oliveira2019, ashruf2024, oliveira2025}. Our work demonstrates how different space environments can lead to varying extents of satellite orbital decay. This work also highlights physical properties of satellites which can make them more resilient (or susceptible) to geomagnetic storm induced orbital drag.

\section{Satellite Orbital Decay Precipitated by Geomagnetic Storms}

Geomagnetic storms are primarily driven by CMEs and corotating interaction regions (CIRs). These storms are characterised by a depression of the horizontal magnetic field of the Earth when such interplanetary structures magnetically reconnect with the Earth's magnetosphere \citep{Lakhina2016}. The intensity of these storms are quantified using geomagnetic indices, like the Disturbance Storm Time (Dst) index. CMEs are typically known to drive moderate to severely intense storms, while CIRs drive weak to moderate storms of relatively longer duration \citep{borovsky2006}. The strength and duration of geomagnetic storms have different impacts on satellite orbits, as we demonstrate below. Also, certain physical characteristics like the ballistic coefficient of satellites can influence their resistance to aerodynamic drag.

We use Precise Orbit Determination (POD) data from the Swarm C satellite and thermospheric data retrieved from it to study the impact of geomagnetic storms on its orbit \citep{Olsen2013, vandenIJssel2020}. The total and geomagnetic storm induced orbital decay is computed according to the method proposed by \cite{Chen2012}. In addition, we use simulated satellite orbits (as discussed in \cite{Baruah2024}) -- validated by POD data from Swarm C -- to study the influence of ballistic coefficients on satellite orbits.

\subsection{Impact of Storm Intensity on Satellite Orbit}

\begin{figure}[H]
  \centering
  \includegraphics[width=\textwidth]{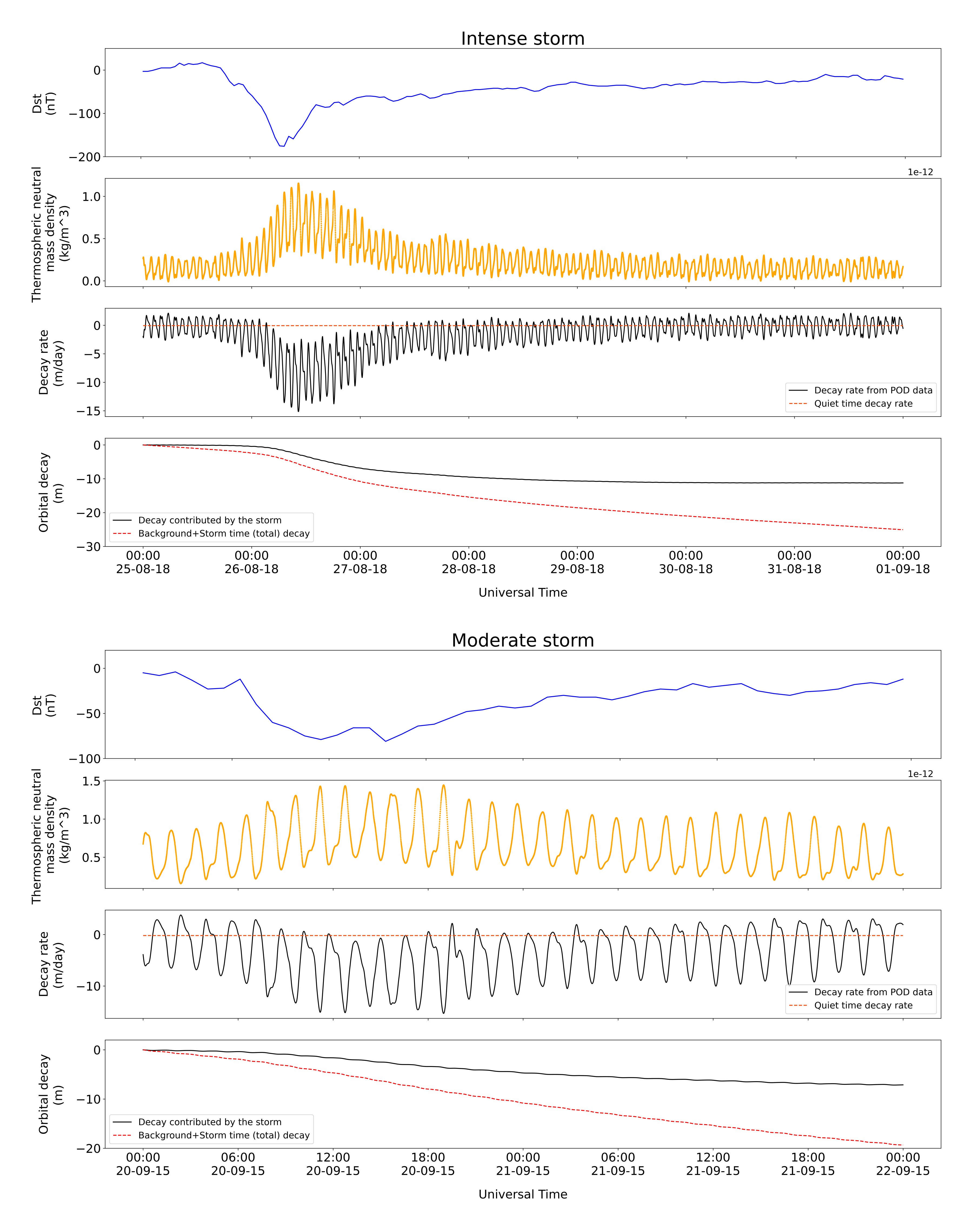}
  \caption{This image shows an analysis depicting the impact of two geomagnetic storms on Swarm C satellite orbit -- an intense storm of 25 August, 2018  and a moderate storm of 20 September, 2015. The plots denote the geomagnetic storm intensity, thermospheric density in the orbit of Swarm C, decay rate, storm-induced orbital decay and the total orbital decay of Swarm C during the two storms.}
  \label{weak_intense}
\end{figure}

We consider two geomagnetic storms of different intensities to study their impacts on the orbit of the Swarm C satellite at an altitude of approximately 455 km, as shown in Fig \ref{weak_intense}. The geomagnetic storm of 25 August, 2018 was an intense storm with a peak Dst of -176 nT. The perturbation caused to the atmosphere by this storm is reflected in the 300\% increase of thermospheric density relative to quiet time conditions in the orbit of Swarm C following the peak of the storm. The decay rate of the satellite increased to a maximum of 17 m/day above quiet time decay rate, resulting in an additional storm-induced orbital decay of 11 m over the duration of the storm. The orbit of the satellite decayed by a total of 25 m during this storm (refer to Fig \ref{weak_intense}).

We next consider a moderate geomagnetic storm of 20 September, 2015 which had a peak Dst value of -81 nT. As a consequence of this storm, the thermospheric density in the orbit of Swarm C increased by 160\%, enhancing the decay rate of the satellite to a maximum of 15 m/day above quiet time decay rate resulting in additional storm induced orbital decay of 6m. The total orbital decay during this storm was 19.35 m.

Thus, it is evident that more intense storms lead to higher loss of altitude of LEO satellites which its to be expected.

\subsection{Impact of CME vs CIR Induced Geomagnetic Storms on Satellite Orbits}

\begin{figure}[H]
  \centering
  \includegraphics[width=\textwidth]{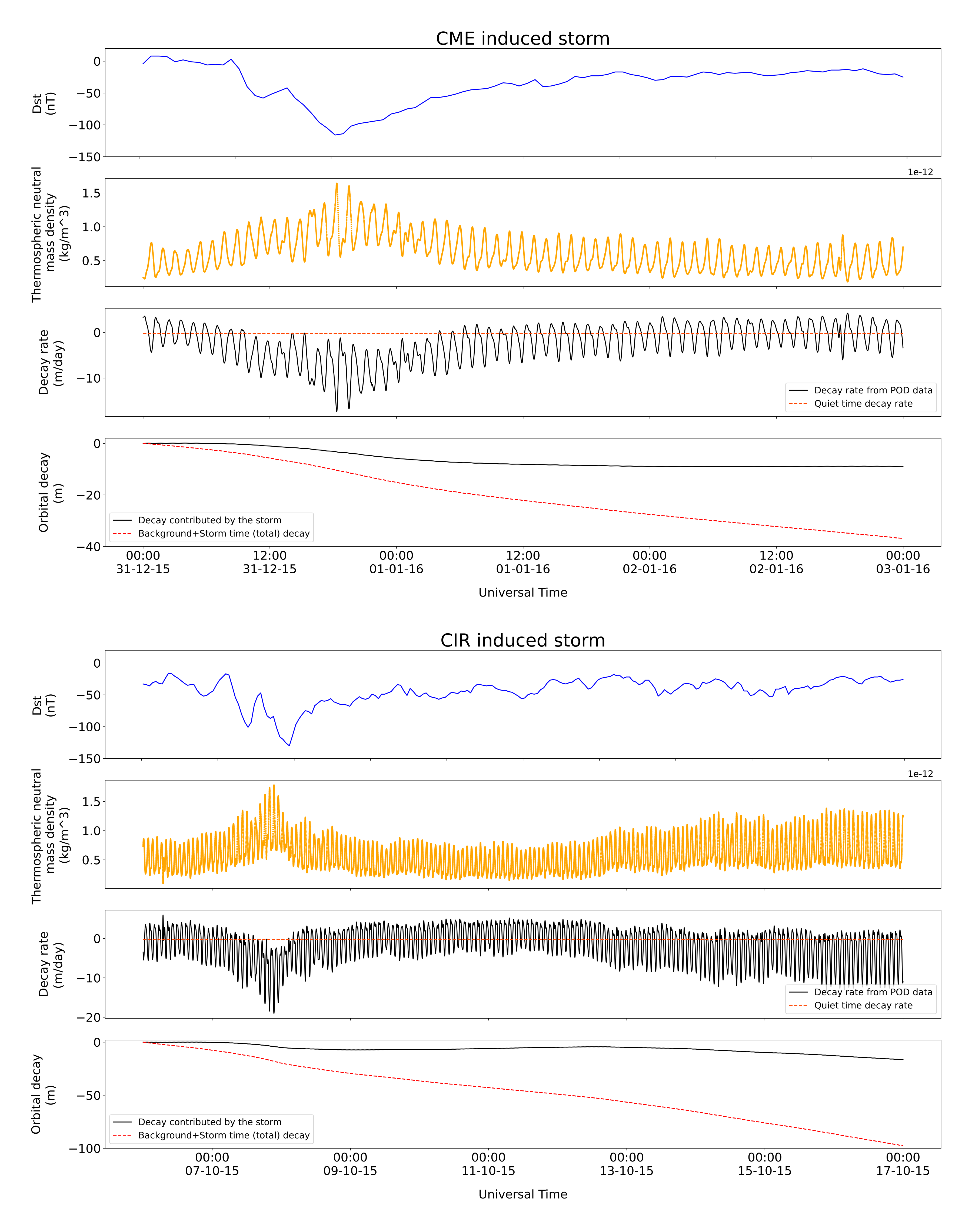}
  \caption{This image shows an analysis depicting the impact of two geomagnetic storms on Swarm C satellite orbit -- a CME induced storm of 31 December, 2015  and a CIR induced storm of 6 October, 2015. The plots denote the geomagnetic storm intensity, thermospheric density in the orbit of Swarm C, decay rate, storm-induced orbital decay and the total orbital decay of Swarm C during the two storms.}
  \label{cme_cir}
\end{figure}

Geomagnetic storms triggered by CMEs and CIRs differ in terms of intensity and duration. CMEs often generate stronger storms than CIRs. However, CIR driven storms persist for a much longer time. In the last section we have already discussed the impact of more intense storms on satellite orbital decay. We now analyse two storms of similar intensity but varying time span -- generated by a CME and a CIR respectively -- and estimate the loss of altitude of Swarm C due to the storms (Fig \ref{cme_cir}).

The moderate geomagnetic storm of 31 December, 2015 was caused by a CME and had a Dst value of -116 nT. As is evident from Fig \ref{cme_cir}, the thermospheric density at the Swarm C altitude was enhanced by a maximum of 400\% relative to quiet times and returned to quiet time values within two days. Swarm C suffered a total orbital decay of 37 m during this storm.

The geomagnetic storm of 6 October 2015 was caused by a CIR with a peak Dst of -128 nT. It enhanced the thermospheric density at $\sim$ 455 km by a maximum of about 400\% but the perturbation persisted for more than 3 days. As a result, Swarm C endured a total orbital decay of 97.6 m during the storm.

We thus conclude that, in case of geomagnetic storms of similar intensity, CIR induced storms can prove to be more detrimental to satellite orbital lifetimes than those induced by CMEs (also suggested by \cite{Wang2021}).

\subsection{Influence of Satellite Ballistic Coefficients on Satellite Orbits}

We use modeled satellite orbits, as discussed in \cite{Baruah2024}, to study the influence of physical properties of satellites on their orbital decay. We first simulate the orbit of Swarm C during the geomagnetic storm of 20 September 2015 (also considered in Section 2.1) and validate it using observations. The estimated orbital decay of Swarm C at 455 km during the time period is 17.28 m, which is in good agreement with the observed orbital decay of 19.35 m. We now use the modelled orbits to simulate the orbital decay of a satellite similar to the International Space Station (ISS) at the same altitude 455 km during the same time frame, under the same geomagnetic conditions. The orbital decay of the ISS like satellite is estimated to be 54.44 m.

This enhanced decay of the ISS-like satellite is attributed to the ballistic coefficient -- $\frac{m}{C_{D}A}$, where m is the mass of the satellite, A is the cross-sectional area and $C_{D}$ is the drag coefficient, for which we assume a generic value of 2.2 in our study. A satellite with a lower ballistic coefficient is subjected to higher aerodynamic drag. The ballistic coefficient of Swarm C is 303.9 kg/m$^{2}$ while that of an ISS-like satellite is 100 kg/m$^{2}$, which explains a higher orbital decay for the latter.

\section{Conclusion}

In this work, we investigate the influence of geomagnetic storms on satellite orbital lifetimes. We conclude that stronger storms induce a higher loss of satellite altitude. For storms of comparable intensity, CIR induced geomagnetic storms tend to result in higher satellite orbital decay than CME induced storms, since CIRs generate sustained geomagnetic storms and perturb the Earth's atmosphere for a longer duration of time \citep{Baruah2024, Roy2023}. Physical properties of satellites, namely the ballistic coefficient also have an impact on their orbital lifetimes. Satellites with lower ballistic coefficients are less resilient to drag and suffer higher orbital decays. Our work demostrates the direct impact of space weather on our space assets and underscores the importance of accurate space weather assessments and predictions to mitigate potential damage to our technological infrastructure.

\section{Acknowledgment}
The Center of Excellence in Space Sciences India is funded by the Ministry of Human Resource Development, Government of India. The author thanks Professor Dibyendu Nandy, CESSI, IISER Kolkata for his valuable feedback. 

\bibliography{iauguide}
\bibliographystyle{iaulike}

\end{document}